\documentclass{article}

\usepackage[OT2, T1]{fontenc}
\usepackage[english]{babel}
\usepackage{graphicx}
\usepackage{authblk}
\usepackage[lmargin=3cm,rmargin=3cm,tmargin=2cm,bmargin=2cm,marginpar=2cm ,reversemp]{geometry}
\usepackage{amsmath,amssymb,bm,color,mathrsfs,wasysym}
\usepackage{hyperref}

\date{}

\title{Inter-Event Time Power Laws in Heterogeneous Systems}
\author[1]{Federico Ettori
\thanks{federico.ettori@polimi.it}}
\author[1,2]{Timothy J. Sluckin
\thanks{t.j.sluckin@soton.ac.uk}}
\author[1]{Paolo Biscari
\thanks{paolo.biscari@polimi.it}}

\affil[1]{\small Department of Physics, Politecnico di Milano, Piazza Leonardo da Vinci 32, 20133 Milan, Italy}
\affil[2]{\small School of Mathematical Sciences, University of Southampton, University Road, Highfield, Southampton, SO17 1BJ, UK}
 
\begin{document}

\maketitle

\begin{abstract}
We investigate the dynamic behaviour of spin reversal events in the dilute Ising model, focusing on the influence of static disorder introduced by pinned spins. Our Monte Carlo simulations reveal that in a homogeneous, defect-free system, the inter-event time (IET) between local spin flips follows an exponential distribution, characteristic of Poissonian processes. However, in heterogeneous systems where defects are present, we observe a significant departure from this behaviour. At high temperatures, the IET exhibits a power-law distribution resulting from the interplay of spins located in varying potential environments, where defect density influences reversal probabilities. At low temperatures, all site classes converge to a unique power-law distribution, regardless of their potential, leading to distinct critical exponents for the high- and low-temperature regimes. This transition from exponential to power-law behaviour underscores the critical response features of magnetic systems with defects, suggesting analogies to glassy dynamics. Our findings highlight the complex mechanisms governing spin dynamics in disordered systems, with implications for understanding the universal aspects of relaxation in glassy materials.
\end{abstract}

\section{Introduction}

The concepts of \textit{relaxation time} $\tau$ and \textit{relaxation length} $\xi$ are ubiquitous in physical systems.  Intuitively, a system ``forgets'' fluctuations on a time scale $\tau$ and length scale $\xi$, and the analogous correlation functions follow exponential decaying laws $ \sim  {\rm e}^{-t/\tau}$ and ${\rm e}^{-r/\xi}$ in time and space respectively.  In the spatial case, this exponential decay follows very generally e.g.\ from the Landau square gradient term in a free energy (see e.g.\ \cite{hohenberg2015,Selinger2016}), while the exponential time decay follows from the first order relaxation equations as expressed long ago e.g.\ by  Langevin \cite{langevin1908} and Onsager \cite{onsager1931}.

However, despite the widespread applicability of the ``exponential decay'' paradigm, it lacks complete universality.  Anomalous temporal  decay   can either be of the so-called ``stretched exponential'' form ($\sim {\rm e}^{-(t/\tau)^\beta}$, with $0<\beta<1$) (e.g.\ \cite{potuzak2011}), or  more dramatically,  algebraic (or power law), taking  the form $\left(\tau/t\right)^{\lambda}$ (e.g. \cite{rieger1993}). Anomalous spatial decay is analogous, although in any given system, the relationship between anomalous spatial and temporal behavior is possible but not obligate.

In fact, power-law behavior seems to emerge when analyzing more complex systems, often including significant spatial heterogeneity.  It  may not only characterize correlation and attenuation, but also the frequency distribution of distinct ``events''. Contexts in which power law distributions are observed include earthquake magnitudes -- the so-called Gutenberg-Richter law (see e.g.\ \cite{gutenberg1949, meng2019}), avalanches in martensitic solids and sand piles  (e.g.\ \cite{hwa1992,xavier2015,zanzottera2016}), Barkhausen noise associated with magnetic switching  \cite{Sethna2001,Puppin2000,Ettori2023}, glasses \cite{potuzak2011}, fragmentation \cite{katsu,wittel}, and dislocation propagation \cite{dimiduk,ispano}.

Given the rather varied contexts in which they occur, power-law distributions have been the subject of extensive study. The question as to the common factors in different physical systems exhibiting similar power-law distributions and correlation remains tantalizingly  open. However,  among mechanisms which have been identified as potential sources of such behavior are percolation, fragmentation, frozen disorder, random walks, criticality, and self-organized criticality. We remark, however, that power-law distributions are found both in the real world (e.g.\ \cite{potuzak2011}) as well as in simulated  simplified models  (e.g.\ \cite{rieger1993}), where we cite only  literature from studies of glassy systems with frozen disorder. This letter reports work in the latter class.

A crucial concept in the analysis of complex systems is the study of the \textit{waiting time} between consecutive events, also known as inter-event time (IET). This idea was borrowed from applied probability, and specifically queuing theory \cite{Kleinrock1975}, and is now widely used with applications ranging from telecommunications to healthcare systems. The key events in the queue are the individual arrivals at the end of the queue. Informally, the mathematics lies in the processing of the queue, and the willingness of the client to engage in queuing given the nature of the queue and the most recent arrival.

If the occurrence of events is Poissonian, i.e.\ independent of the occurrence of any previous event, then the IET pdf is exponential, corresponding to exponential decay of time correlations. When, for whatever reason, the probability of an event occurring is affected by the proximity of a recent event, then the probabilities that  two neighboring events take place a time $t$ apart are no longer independent. The resulting so-called ``fat-tailed'' distributions decay more slowly than exponentially.

There are numerous practical examples from the human sciences in which event occurrence is no longer Poissonian  \cite{Barabasi2005,Oliveira2005,Vazquez2006,Crane2010}.  Earthquake frequencies \cite{gutenberg1949, meng2019} exhibit similar fat-tailed behavior.  However,  at first sight, geophysical and anthropological systems seem to offer few obvious analogies, other than the large number of interacting degrees of freedom and the heterogeneity of the interactions.

The capacity of surprisingly simple models to illuminate complex behavior has been much noted by historians of science. In the context of magnetism, the Ising model, involving interactions between nearest neighbor classical spins in  a $d$-dimensional lattice,  will celebrate its 100th birthday in 2025, but remains remarkable vigorous. The Edwards-Anderson spin glass \cite{edwards1975}, involving random bond interactions on a lattice of spins, is only approaching its 50th birthday, but is likewise the subject of continuing attention. The related random field spin model \cite{imry1975, chalker1983,fytas2013,ding2024} is of similar vintage, also seems to exhibit glassy behavior. Variants of this have proved fruitful and challenging to mathematicians and physicists alike, as in random anisotropy magnets where numerical simulations of the dynamic behavior revealed commonalities with glassy-like systems(e.g.\ \cite{Garanin2024,Garanin2022}).  In this Letter, we  study  the dilute Ising model with pinned spins, some of whose static and dynamic properties we have previously investigated \cite{ettori2023b, Ettori2024}.

As discussed in \cite{ettori2023b}, certain large-scale properties of the present model are analogous to those of the Random Field Ising model (RFIM). Specifically, the model with defects can be interpreted as a specific case of the RFIM, where fixing spins corresponds to introducing strong local fields with random orientations at the random defect sites. Both models exhibit similar equilibrium properties, including cluster formation in the thermodynamic limit and spontaneous magnetization in finite-sized systems. These \emph{cluster} phases are a particularly compelling research topic, as they possess equilibrium properties that are intermediate between paramagnetic phases (characterized by zero net magnetization and the existence of a unique free energy basin) and the spin-glass phases (defined by the absence of long-range order, nonzero local average magnetization). While a more detailed characterization of these equilibrium properties remains an active area of research, this study focuses on their out-of-equilibrium dynamics.

The key prerequisite is the existence of a dynamic phase transition in spin models. The static ferromagnetic transition involves the  development of spontaneous order, and hence the inability of the system to follow all external ordering fields.  Suppose that, instead of imposing a static external field, one imposes an ordering field periodic in time, i.e.\ the ordering field is dynamic.

In a pure Ising system, under suitable phase conditions (usually high temperatures) the spins follow the ordering field, with the same frequency as the ordering field. This is the Dynamic Disordered Phase (DDP). However, for low enough temperatures, and low amplitude fields, a Dynamic Ordered Phase (DOP) emerges. In this phase the spins have preferred one or other of the directions of a local molecular field. The spins no longer, on average, follow the imposed external field. The Dynamic Phase Transition (DPT) marks the boundary between the two types of behavior \cite{Tome1990,jung1990,sides1998,Ettori2024}. We refer to reference \cite{Yuksel2022} for an extensive review of the topic. Although the spins no longer follow the oscillating field completely, local spin flips do still occur. The distribution of spin flip events can be investigated.

We find here that in a pure Ising system these spin flips seem to occur in Poissonian fashion and hence that the lifetime distribution is exponential. We also investigate the analogous problem in the dilute Ising model with pinned spins. As in the case of the static phase transition, the character of the dynamic phase transition changes discontinuously with the addition of the pinned sites. There is still a disordered phase. But now the system is heterogeneous;  different sites behave in different ways.  In the ordered phase, some sites are able to follow the oscillating field, and some are not. Of those that are not, some point preferentially in a $+$ direction and some in a $-$ direction.  The spin flip events still occur,  from directions which are locally ``favored'' into regions which are locally disfavored, although they are increasingly unlikely.  The pdf of the inter-event times acquires a fat tail, and changes from exponential to algebraic. This is our principal result.

\section{Model} The system under investigation is the 2D nearest-neighbour Ising model, in which a fraction of randomly chosen spins (defects) are held fixed in both orientation and position throughout the entire simulation. To maintain neutrality, half of the defects are positive ($s_i=+1$) and the other half are negative ($s_i=-1$). We investigate the dynamic response of this system to a sinusoidal, time-dependent magnetic field, $h(t) = h_0 \cos\omega t$. At a critical temperature $\Theta_\textrm{\footnotesize c}$ or, alternatively, a critical frequency $\omega_\textrm{\footnotesize c}$ or field amplitude $h_{0,\textrm{\footnotesize c}}$, the manner in which the system responds changes.

For $T > \Theta_\textrm{\footnotesize c}$ (or $\omega < \omega_\textrm{\footnotesize c}$, or $h_0 > h_{0,\textrm{\footnotesize c}}$), the system enters in the DDP, in which the time-dependent magnetization follows the external field, possibly with a time delay. Conversely, for $T < \Theta_\textrm{\footnotesize c}$ (or $\omega > \omega_\textrm{\footnotesize c}$, or $h_0 < h_{0,\textrm{\footnotesize c}}$), the system enters in the DOP, in which the magnetization remains trapped in a low-energy state and reversal events become rarer. The order parameter of the DPT is the average magnetization per cycle $Q = \omega/2\pi\oint m(t) \,dt$. As for the equilibrium phase transition, the critical temperature is associated with a peak in the susceptibility of the order parameter: $\chi_Q = L^2(\langle Q^2\rangle-\langle Q\rangle^2)$. In a previous work \cite{Ettori2024}, we discussed how the introduction of defects generally decreases $\Theta_\textrm{\footnotesize c}$. In this work, we focus on the reaction time of individual spins in both the homogeneous (defect-free) case and the heterogeneous (with defects) case.

To identify true reversals among numerous fluctuation events, we track the time-averaged value of each individual spin $s_i(t)$ over a half-cycle of the external field, expressed as
\begin{equation}
	\overline{s_i(t)}=\frac2P \int_{nP/2}^{(n+1)P/2} s_i(t)dt,
\end{equation}
where $P$ is the period of the external field. A reversal is considered to occur in a specific half-cycle when the time average changes sign. The local reversal time $n_i$ is defined as the number of external field half-cycles between two consecutive reversals. This metric captures the system's temporal response. Short average local reversal times correspond to systems which follow the external field more easily, while longer reversal times are observed when the spins tend to retain their orientation over time, despite the oscillations of the field.

We define the reversal time probability distribution $P_\textrm{\footnotesize rev}(n)$ by calculating the local reversal times for all free spins in the lattice. Additionally, we determine the avalanche probability distribution $P_\textrm{\footnotesize aval}(S)$, which represents the likelihood of observing a connected cluster of $S$ sites undergoing simultaneous reversal events, where by \emph{simultaneous} we mean occurring during the same external field half-period.

The system is analyzed using Monte Carlo simulations with the $N$-Fold way algorithm \cite{Bortz1975} and Glauber dynamics \cite{Glauber1963}.
This study focuses on low temperatures, where the $N$-Fold way algorithm is more computationally efficient than the more common traditional Metropolis algorithm \cite{Bortz1975}. The Glauber dynamics is preferred for dynamic studies, as it more accurately represents a true dynamical process than the Metropolis algorithm \cite{Ito1997}, which was invented to converge in the equilibrium limit to a Boltzmann distribution. For all simulations, a lattice size of $L=200$ and a period $P=2\pi/\omega=258$ were used. Henceforth, the time unit is set equal to the characteristic time associated with the Glauber transition probability rates, while temperature and magnetic fields are expressed in units of $J/k_B$ and $J$ respectively, where $k_B$ is the Boltzmann constant and $J$ is the Ising exchange interaction constant.

\section{Homogeneous systems} 
\begin{figure}
	\includegraphics[width=8.0cm]{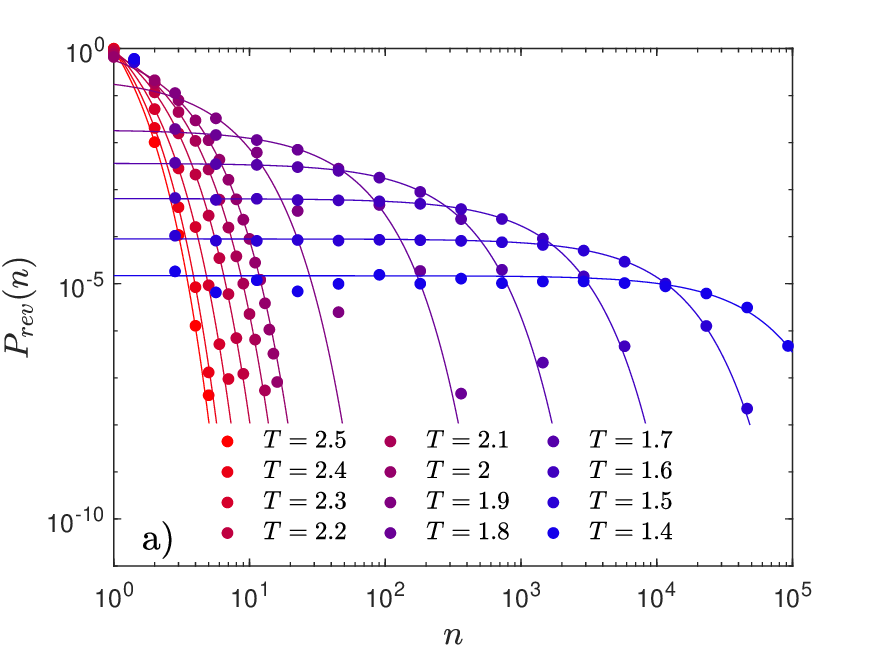}
	\includegraphics[width=8.0cm]{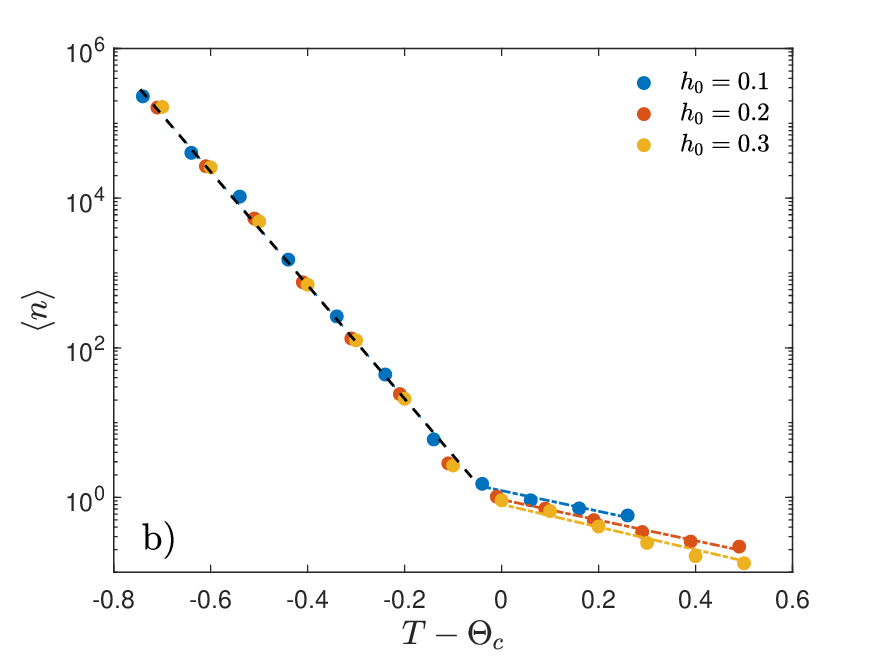}
	\caption{IET probability distribution for the reversal time in a homogeneous system. Panel a): Probability distribution $P_\textrm{\footnotesize rev}(n)$ in a double logarithmic plot for $h_0=0.3$ at different temperatures. Solid lines represent the fit to the most likely exponential distribution, while dots correspond to simulation data. Panel b): Average reversal time as a function of temperature for different field intensities. The error bars derived considering 10 independent simulations are of the same size as the markers and are not shown for clarity. In the low-temperature phase, a single linear fit (dashed line) is applied across all field intensities. In the high-temperature phase, separate fits are performed and shown for each field intensity.}
	\label{fig:PnNoDefects}
\end{figure}

We begin by analyzing the dynamic response of the defect-free Ising model. Fig.~\ref{fig:PnNoDefects}.a shows the IET probability density function $P_\textrm{\footnotesize rev}(n)$ for several temperatures, spanning both the dynamical ordered and disordered regimes. The solid lines in the plot display the best fit to a discrete exponential distribution $P_\textrm{\footnotesize rev}(n) \sim \textrm{\footnotesize e}^{-\lambda n}$, where the characteristic parameter $\lambda$ is determined through the maximum likelihood (ML) method for each temperature and field amplitude. The close match obtained at all temperatures confirms that the local spin-flip process is basically Poissonian.

Fig.~\ref{fig:PnNoDefects}.b shows the dependence of the average reversal time, $\langle n \rangle = \lambda^{-1}$, on both temperature and external field amplitude. The curves corresponding to different field amplitudes nearly collapse onto a single curve when the average reversal time is plotted as a function of the temperature deviation from the (field-dependent) critical point. This collapse is most prominent in the low-temperature phase. Furthermore, in a semi-logarithmic plot, the relationship appears almost linear, indicating that $\langle n \rangle$ depends exponentially on temperature, with two distinct slopes in the ordered and disordered phases, and a crossover at the critical point, where $\langle n \rangle\approx 1$.

\section{Heterogeneous systems} 
\begin{figure}
	\centering
	\includegraphics[width=7.5cm]{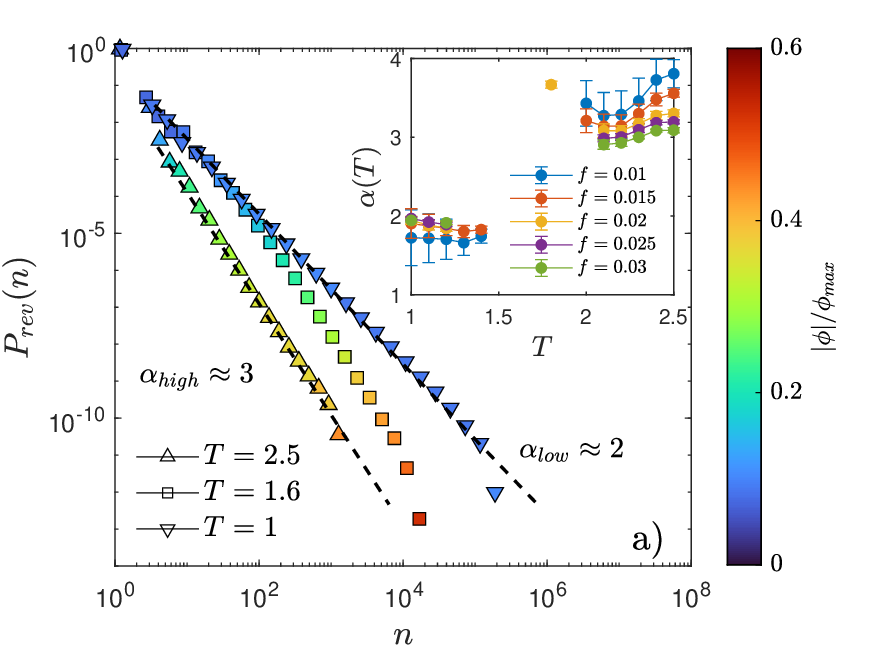}
	\includegraphics[width=7.5cm]{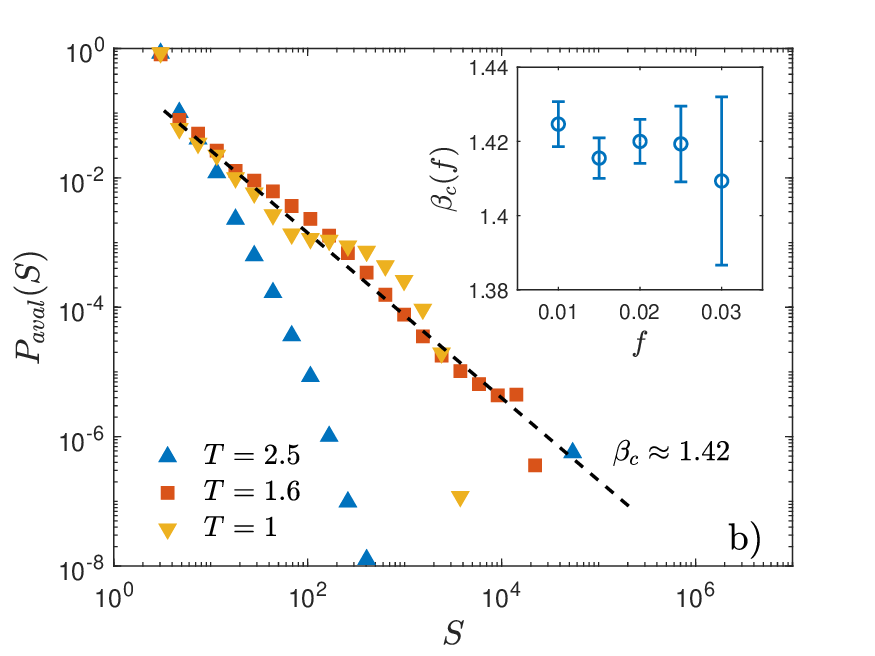}
	\caption{Probability distribution of a) the IET, and b) avalanche size, for heterogeneous systems with $h_0=0.3$, $f=0.025$, and $T=2.5, 1.6, 1$ (markers: logarithmic binning of data points; dotted lines: best power-law fit). The colour scale represents the average potential absolute value associated with each measurement bin. Panel a): $P_\textrm{\footnotesize rev}(n)$ exhibits power-law behaviour in both the low- and high-temperature regimes, with distinct exponents ($\alpha_{high}$ and $\alpha_{low}$). The inset shows the ML estimation of the power-law exponent as a function of temperature for different defect fractions. Data points are shown where the GoF test returns a $p$-value greater than 0.1. Panel b): $P_\textrm{\footnotesize aval}(S)$ is dominated by large clusters in the high-temperature regime, while it follows a power-law distribution for $T = \Theta_\textrm{\footnotesize c}$. In the low-temperature regime, exponential damping is observed. The inset presents the ML estimation of the critical exponent as a function of the defect fraction.}
	\label{fig:PnDefects}
\end{figure}

We now analyze the inter-event times (IETs) in an Ising model with defects, subject to an oscillating external field. Our primary finding is that, away from criticality, a collective behavior emerges, where $P_\textrm{\footnotesize rev}(n)$ follows a power-law distribution over several orders of magnitude (up to 6). We will examine this distribution, focusing on the power-law exponents and their dependence on temperature and defect fraction. Fig.~\ref{fig:PnDefects}.a shows $P_\textrm{\footnotesize rev}(n)$ for a defect fraction of $f = 0.025$ and field amplitude $h_0 = 0.3$. At both high and low temperatures, $P_\textrm{\footnotesize rev}(n)$ can be approximated by a discrete power-law distribution. The power-law exponents are estimated using the ML method, we validate the power-law hypothesis using a goodness-of-fit (GoF) test \cite{Clauset2009} over an intermediate range of values for $n$ (excluding the eventual exponential cutoff), and accept the hypothesis when the $p$-value exceeds 0.1. We slightly relax the test and require, for a positive outcome, that the Kolmogorov-Smirnov distance of the data is at maximum 10 times larger than the one from synthetic data. The inset of Fig.~\ref{fig:PnDefects}.a shows the power-law exponents, which are approximately 2 in the ordered (low-temperature) phase and 3 in the disordered (high-temperature) phase.

The intermediate distribution shown in Fig.~\ref{fig:PnDefects}.a, corresponding to the critical temperature, does not meet the criterion for a power-law distribution according to the GoF test, so no exponent is reported in the inset for this and similar temperatures. Additionally, the inset highlights that the power-law exponents do not show a significant dependence on defect fraction $f$, at least within the 1–3\% range of defects. Similarly, no significant variation in the critical exponents is observed with changes in system size. The colour coding in this plot will be discussed below when we analyze the defect potential.

In Fig.~\ref{fig:PnDefects}.b, we analyze the avalanche size probability distribution, $P_\textrm{\footnotesize aval}(S)$. For each half-cycle, we identify the sites that have undergone reversal events and group them into connected clusters of simultaneously reversing sites. We then study the cluster size probability distribution across different temperatures and defect fractions. Near the critical temperature (red squares), we observe that avalanche sizes follow a power-law distribution spanning over 4 orders of magnitude. In the low-temperature phase (yellow downward triangles), a similar power-law emerges, but with a notable cutoff that prevents the formation of very large avalanches. In contrast, in the high-temperature (blue upward triangles), disordered phase, a power-law is not observed. In this case, most events involve large, connected sets of spins reversing in response to the external field oscillations.

The inset of Fig.~\ref{fig:PnDefects}.b shows the ML estimate of the power-law exponent at the dynamic critical temperature, $\Theta_\textrm{\footnotesize c}(f)$. Within the error bars, the critical exponent shows no significant dependence on defect fraction. Notably, before the exponential cutoff, the low-temperature cluster size distribution follows a power-law with the same critical exponent measured at the critical temperature—a phenomenon similar to the high-temperature Barkhausen noise cutoff reported in \cite{Ettori2023}.

\section{Defect potential} The defect-induced spatial symmetry breaking strongly influences the spin reversal process. It is the aim of the present analysis to understand whether it is possible to predict and quantify how different local sites are expected to behave when an oscillating field is applied on the system, with particular attention focused on the IET probability distributions discussed in the above section.

We use the defect potential introduced in \cite{Ettori2024}
\begin{equation}
	\phi_i = \sum_j \frac{q_j}{d(i,j)},\label{defpot}
\end{equation}
where $q_j$ is the defect's \emph{charge} (i.e., spin value) at position $j$, and $d(i,j)$ is the Manhattan distance consistent with periodic boundary conditions) between the free site $i$ and defect $j$. As demonstrated in \cite{Ettori2024}, the defect potential provides a useful tool for determining and quantifying which sites are more/less affected by the quenched defects. In the sec.\ \ref{sec:SM}, we show that this is confirmed also when we study the DPT, as several quantities are strongly correlated with the defect potential, including the average magnetization, the local magnetization per cycle, the local reversal probability, and the local average inter-event time.

\begin{figure}
	\centering
	\includegraphics[width=\linewidth]{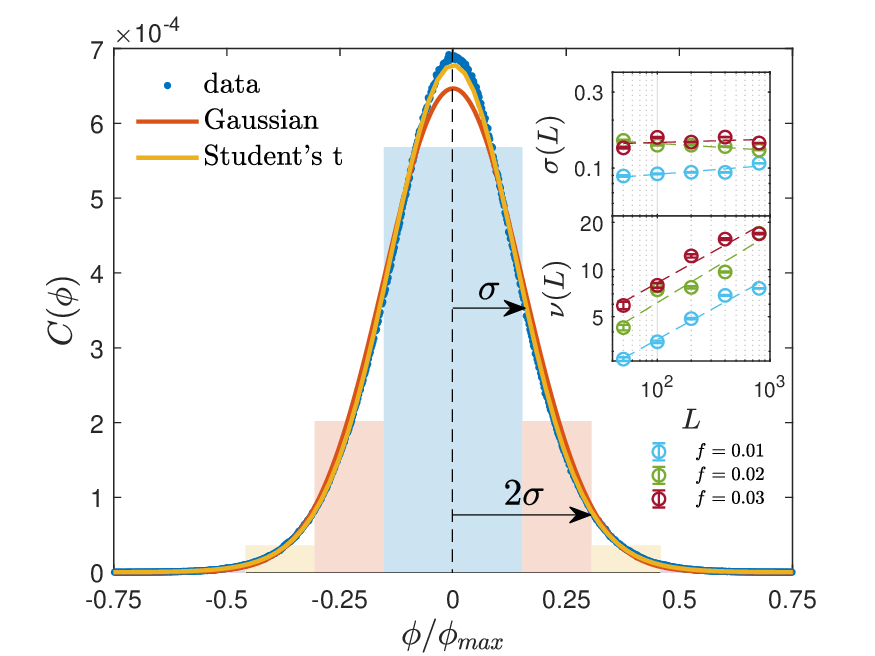}
	\caption{Defect potential distribution for a square network of size $L=800$ with a quenched defect fraction $f=0.03$. The fitted curves suggest that the distribution is optimally approximated by a Student's $t$. The insets illustrate how the standard deviation and the parameter $\nu$ depend on the defect fraction and the system size. The blue box spans a width of $2\sigma$, with a height proportional to the fraction of network sites exhibiting defect potentials below $\sigma$ in absolute value. The height of the red box represents the proportion of sites with defect potentials between $\sigma$ and $2\sigma$, while the yellow boxes extend to the $3\sigma$ threshold.}
	\label{fig:distr}
\end{figure}

Our goal is to determine whether sites with high and low values of the defect potential contribute similarly or differently to the IET probability distribution shown in Fig.~\ref{fig:PnDefects}.a. To this end, we first analyzed how the distribution of defect potentials depends on the system size and defect fraction. Overall defect neutrality implies the distribution is symmetric, with potential values $\phi$ and $-\phi$ having the same probability of occurrence. Fig.~\ref{fig:distr} presents the distribution of potential values for a square network of $800 \times 800$ spins with a fraction $f=0.03$ of quenched defects. The fitted curves indicate that this distribution is well-approximated by a Student's \(t\)-distribution, characterized by a standard deviation $\sigma$ and parameter $\nu$. The inset shows how these two parameters depend on the system size and the defect fraction. The relative standard deviation $\sigma/\phi_{max}$ is constant over the system size for all the considered defect fractions. The divergence of the Student's $t$ parameter $\nu$ with the system size confirms that, consistent with the Central Limit Theorem, in the thermodynamic limit the defect potential distribution approaches a Gaussian shape. 

We compute the standard deviation $\sigma$ of the defect potential distribution and classified the sites into categories: low (sites having $|\phi|$ within $\sigma$), intermediate (larger, but within $2\sigma$), high (even larger, but within $3\sigma$), and very-high (above $3\sigma$) defect potential. Clearly, each class was less populated than the preceding one.

\begin{figure}
	\centering
	\includegraphics[width=7.5cm]{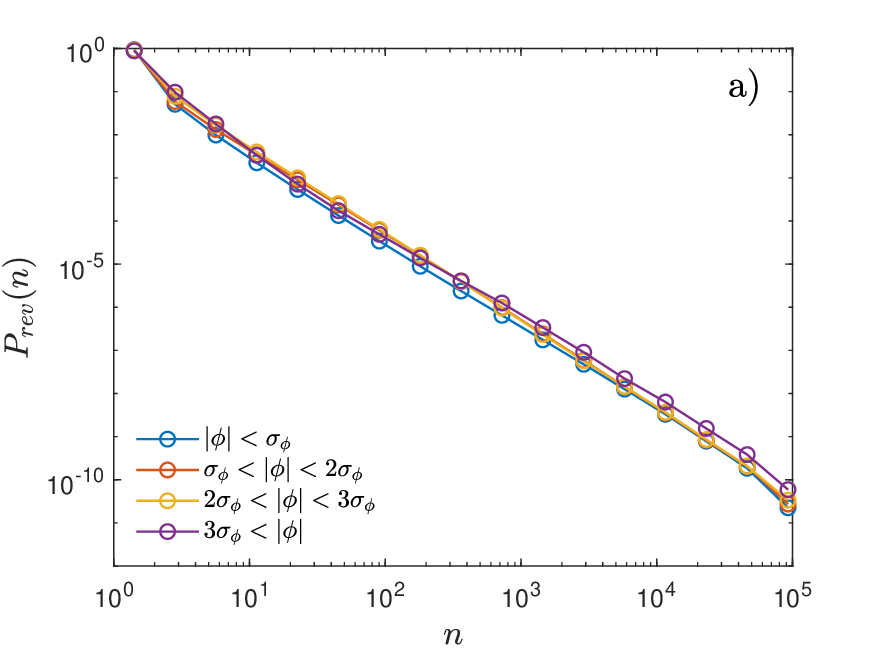}
	\includegraphics[width=7.5cm]{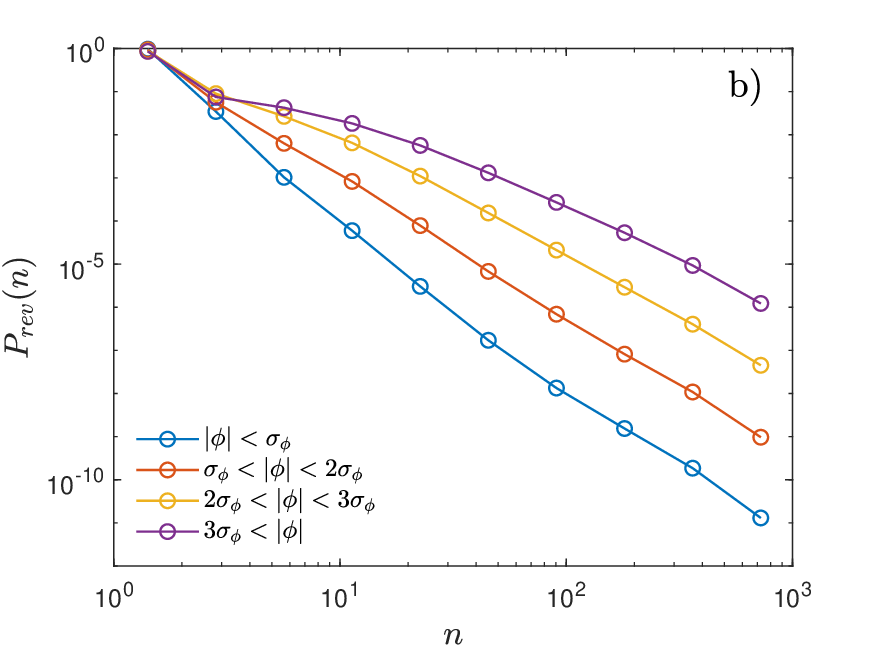}
	\caption{Decomposition of the probability distributions $P_\textrm{\footnotesize rev}$ plotted in Fig.~\ref{fig:PnDefects}.a by considering separately sites having similar values of the defect potential. Panel a): The low-temperature decomposition shows that all sites obey one and the same power-law distribution. Panel b): The high-temperature decomposition shows that different sites behave differently, with the high-potential sites exhibiting distributions with fatter tails.}
	\label{fig:decomposition}
\end{figure}

Fig.~\ref{fig:decomposition} shows how the four different classes contribute to the high- and low-temperature IETs in Fig.~\ref{fig:PnDefects}.a. The most notable result is that in the low-temperature regime (panel a) all classes follow the same power law, which clearly coincides with the overall probability distribution shown in Fig.~\ref{fig:PnDefects}.a. This indicates a remarkable universality that applies to all sites, regardless of their position relative to the quenched defects. Since the number of sites in the first class ($|\phi|<\sigma$) is significantly larger than that in the other classes, the average defect potential $\overline{|\phi(n)|}$ for all sites undergoing reversals after $n$ cycles is dominated by the low-potential sites. This explains why the high-temperature curve in Fig.~\ref{fig:PnDefects}.a is consistently blue, which is the colour representing low potential in the displayed colour scheme.

The situation changes in the high-temperature regime, as shown in Fig.~\ref{fig:decomposition}.b. The IETs of the low-potential sites (blue circles) still obey a power-law distribution, but sites with higher potential values exhibit distributions with significantly fatter tails. The remarkable result here is that the combination of non-power-law distributions from different site classes adds up to produce the overall high-temperature power-law distribution seen in Fig.~\ref{fig:PnDefects}.a. The fact that low-potential sites are much less likely to exhibit long reversal times explains the interesting colour pattern in the high-temperature distribution of Fig.~\ref{fig:PnDefects}.a. Specifically, sites associated with longer reversal times are also associated with high defect potentials. Consequently, the distribution exhibits a pronounced colour shift.

\subsection{Local properties and Defect potential} \label{sec:SM}

\begin{figure}
	\centering
	\includegraphics[width=\linewidth]{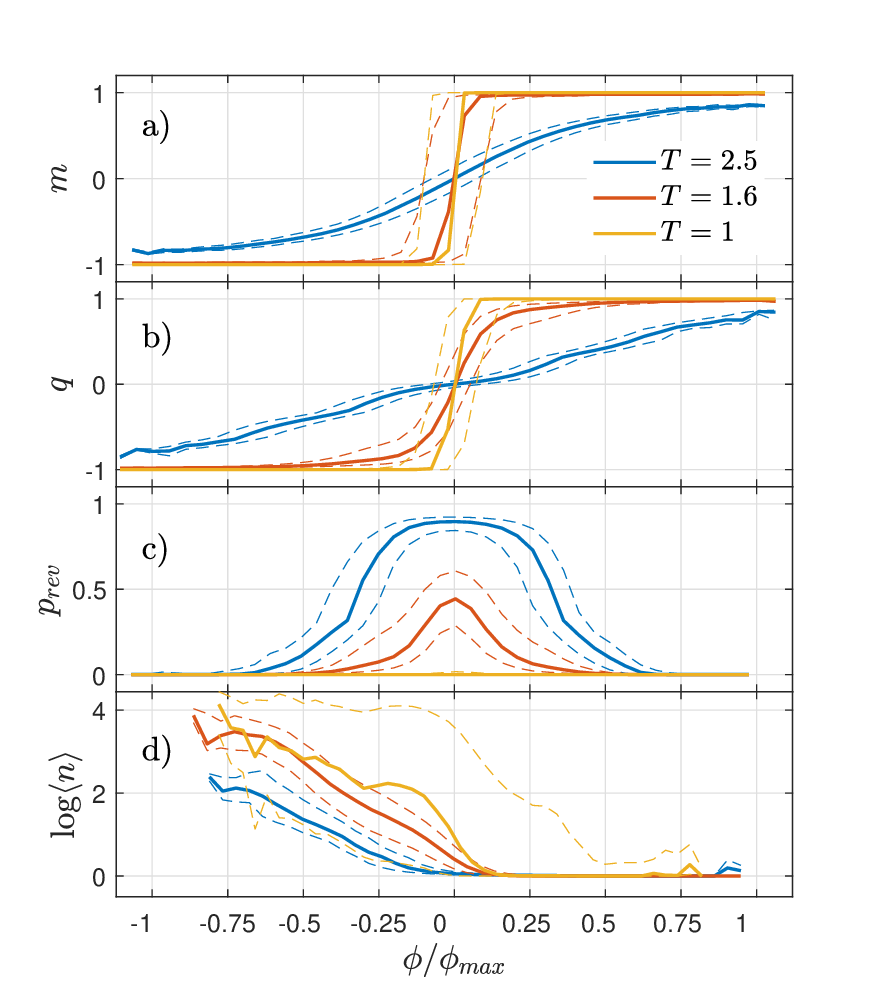}
	\caption{
		Influence of the local defect potential $\phi$ (normalised to its maximum value) and the local properties of the Ising model with defects with $h_0=0.3$ (for panel b, c, d) and $f=0.025$ at different temperatures. The solid lines represent the median value whereas the dotted lines provide information about the range of values which encompass the second and third quartiles of the sites with the reported potential.\\
		Panel a): average spin magnetization at equilibrium. Panel b): local average magnetization per cycle. Panel c): local reversal event probability. Panel d): local average inter-event time.}
	\label{fig:Phi}
\end{figure}

This study investigates the dynamic response of systems with quenched defects under an oscillating external field. Fig.~\ref{fig:Phi} presents the correlation between key physical quantities associated with this response and the defect potential. Simulations over a spin network of side length $L=200$ were performed for all the reported data. In all plots, solid lines indicate the median values across all sites with the same defect potential. Dotted lines depict the range containing the two intermediate quartiles of sites sharing the same potential, providing insight into the dispersion of the physical quantities.

Figure~\ref{fig:Phi}.a shows the local equilibrium magnetization, $ m_i = \langle s_i \rangle $, as a function of the defect potential, $\phi_i$, at various temperatures in the absence of an external field. For this case, averages over quenched randomness were obtained using a multi-spin algorithm with adapted parallel tempering. $100$ batches of $64$ replicas were used. Within each batch, the replicas have defects located on the same sites, whereas the defect positions are independent between batches. Notably, the relationship between magnetization and defect potential resembles the external field-magnetization curves observed in para- and ferromagnetic systems. This similarity suggests that the defect potential may be interpreted as a local field acting on each site, drawing an analogy between the Random-Field Ising model and the Ising model with quenched defects.

In the presence of an oscillatory external field, the order parameter used to study the dynamic response is the local average magnetization per cycle [30]
\begin{equation}
	q_i = \frac{1}{P}\int_{nP}^{(n+1)P}\langle s_i(t)\rangle \, dt
\end{equation}
For this and the following panels, we considered the $N$-Fold way algorithm outlined in the main text in the model section. $100$ defect realisations were simulated over 100 cycles of the external field. Fig.~\ref{fig:Phi}.b shows how the dynamic order parameter depends on the defect potential $\phi$. The strong similarities between the two curves $m(\phi)$ and $q(\phi)$ (with a Spearman correlation coefficient greater than 0.6 across all temperatures and defect fractions) suggest that the configuration $s_i = \textrm{sgn}(\phi_i)$ is a promising candidate for the ground state configuration of an Ising spin system with defects, applicable to both dynamic and static conditions.

The panel in Fig.~\ref{fig:Phi}.c shows the probability of a site with a given defect potential value undergoing a reversal event of any sign. Sites with low absolute values of the defect potential are significantly more likely to experience reversals. Notably, at the dynamic phase transition, the peak probability value (which corresponds to the reversal probability of a low-potential site) equals $\frac{1}{2}$. 

Finally, Fig.~\ref{fig:Phi}.d shows the average inter-event time $\langle n \rangle$ as a function of the defect potential. For this average, only reversals from negative to positive magnetization are considered. This selection explains why the inter-event time is much longer for sites with a negative defect potential, as these sites tend to retain their negative magnetization and require significantly more time to undergo a reversal. The influence of the potential is less pronounced above the critical temperature, although sites with large negative potential consistently require many cycles before experiencing a reversal.

\section{Discussion}

We have performed Monte Carlo simulations of the Ising model under both homogeneous (defect-free) and heterogeneous (with defects) conditions. Our results reveal notable features in the system's dynamic response. Specifically, the inter-event time (IET) between local reversal events follows an exponential distribution in the homogeneous system, while two different power-law distributions emerge in the high- and low-temperature phases of the heterogeneous system, where defects introduce quenched randomness.

Interestingly, in the heterogeneous system, the high- and low-temperature regimes reveal distinct mechanisms that give rise to power-law distributions. At high temperatures, $P_\textrm{\footnotesize rev}(n)$ results from the combined contributions of sites at different potential levels. Generally, spins located far from defects or in regions with balanced opposing defects (low-potential sites) follow the external field more readily, resulting in shorter reversal times. In contrast, spins in defect-dense areas (high-potential sites) are more resistant to switching, contributing to the tail of the probability density function. At low temperatures, all site classes exhibit a similar power-law distribution of IETs, regardless of potential value. These two distinct mechanisms lead to the formation of power-law distributions with different critical exponents, $\alpha_\textrm{\footnotesize high}$ and $\alpha_\textrm{\footnotesize low}$, in the high- and low-temperature regimes, respectively.

Our analysis supports the suggestion that magnetic systems with defects exhibit critical response features peculiar of glassy systems \cite{2020SG, 2022PRX}. More precisely, in pure (non-defected) systems the spin magnetization reversal follows a classical Poissonian waiting-time process leading to exponential inter-event distributions. On the contrary, in the presence of defects, the reversal process induced by an external field is characterized by a critical power-law distribution. Moreover, significant differences appear between the dynamically ordered and the dynamically disordered phase, the latter being characterized by a peculiar universal power-law for all sites (evidenced in Fig.~\ref{fig:decomposition}.a).

\bibliography{MyBibliography_short}
\bibliographystyle{elsarticle-num}

\end{document}